\documentclass[twocolumn,aps,pra,superscriptaddress,showpacs,tightenlines]{revtex4}
\usepackage[T1]{fontenc}
\usepackage[latin9]{inputenc}
\usepackage{amsmath}
\usepackage{graphicx}
\usepackage{amssymb}

\begin{document}

\title{M\"{o}bius Graphene Strip as Topological Insulator}
\author{Z. L. Guo}
\affiliation{School of Physics, Peking University, Beijing, 100871, China}
\author{Z. R. Gong}
\affiliation{Institute of Theoretical Physics, The Chinese Academy of Sciences, Beijing,
100080, China}
\author{H. Dong}
\affiliation{Institute of Theoretical Physics, The Chinese Academy of Sciences, Beijing,
100080, China}
\author{C. P. Sun}
\affiliation{Institute of Theoretical Physics, The Chinese Academy of Sciences, Beijing,
100080, China}
\date{\today}

\begin{abstract}
We study the electronic properties of M\"{o}bius graphene strip with
a zigzag edge. We show that such graphene strip behaves as a
topological insulator with a gapped bulk and a robust metallic
surface, which enjoys some features due to its nontrivial topology
of the spatial configuration, such as the existence of edge states
and the non-Abelian induced gauge field. We predict that the
topological properties of the M\"{o}bius graphene strip can be
experimentally displayed by the destructive interference in the
transmission spectrum, and the robustness of edge states under
certain perturbations.
\end{abstract}

\maketitle


\section{INTRODUCTION}

Because of the unusual properties and potential applications, topological
insulators have recently been under great focus both experimentally~\cite%
{topo01,topo02} and theoretically~\cite{topo1,topo2,topo3,topo4,topo5,topo6,topo7}%
. The topological insulator system belongs to a novel category, possessing
an insulating bulk, with a gap in the energy spectrum of propagating
electrons, whereas its surface is metallic. The edge states promise such
metallic feature and describe the electrons localized on the surface with
energy levels lying just within the gap of the bulk~\cite{topo1}. For many
systems in this category, the surfaces are no longer conductive when some
perturbations are applied. However, for some system with nontrivial
topology, the edge states are robust under perturbations. Such systems with
robust metallic surfaces are referred to as topological insulators~\cite%
{topo1}.

It is natural to imagine that those topological features of
electrons can be realized through nontrivial topology in the
configuration space of the system considered. A most recent
illustration is the tight binding model for electrons hopping on a
M\"{o}bius ladder~\cite{Moladder1}. In this investigation,
observable effects of M\"{o}bius boundary condition were found for a
finite lattice. Destructive interference emerges from the
transmission spectrum to display the typical topological feature. It
was proved that such destructive interference can be explained in
terms of a non-Abelian gauge field induced by the nontrivial
topology. However, such a novel nanostructure can not be regarded as
a topological insulator, because this quasi-1D system has no edge
states.

\begin{figure}[ptb]
\begin{centering}
\includegraphics[bb=59 296 487 768,clip,width=3in]{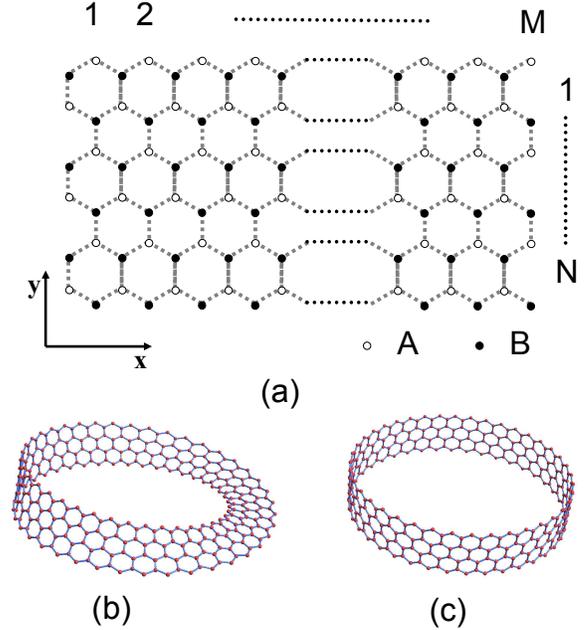}
\par\end{centering}
\par
\centering{} \caption{ (color online) (a) A zigzag graphene strip is
schematically illustrated in a 2D version with open boundary
condition along $y$ direction. The white (black) dots represent
sublattice A (B). The 3D version of the graphene strip with
M\"{o}bius boundary condition along the $x$ direction
($\protect\psi_{a(b)}(\mathbf{r})=\protect\psi_{b(a)}(\mathbf{\widetilde{r}}%
+L\mathbf{e_{x}})$, where $\mathbf{r}=(x,y)$ and $\widetilde{\mathbf{r}}%
=(x,-y) $) and periodic boundary condition ($\protect\psi_{a(b)}(\mathbf{r})=%
\protect\psi_{a(b)}(\mathbf{r}+L\mathbf{e_{x}})$) are plotted in (b) and
(c), respectively.}
\label{fig1}
\end{figure}

In this paper, we study the electronic properties of a M\"{o}bius
graphene strip. We show that with a 2D nontrivial topological
structure, the M\"{o}bius graphene
strip~\cite{Mobius1,Mobius2,Mobius3,Mobius4,Mobius5,Mobius6} with
zigzag edges (see Fig.~\ref{fig1}(b)) behaves as a typical
topological insulator since it possesses robust edge states. It is
noticed that there is no edge state in such M\"{o}bius strip with an
armchair edge, thus no such nontrivial topological properties
appear. Our investigation will focus on the situation with a zigzag
edge. Through analytic approaches, we first predict that the
robustness of the edge states is maintained under perturbation with
uniform electric field in our discussion. And we compare the
M\"{o}bius graphene strip with a generic one whose edges turn to be
insulators in the same electric field. Besides, destructive
interference in the transmission spectrum is found in the M\"{o}bius
graphene strip, which is caused by the non-Abelian gauge field.
Based on this observation, we propose a possible approach of quantum
manipulation on the transport properties of the M\"{o}bius graphene
strip through a magnetic flux.

%


This paper is arranged as follows. In Sec.~\ref{sec:edge}, we give
an analytical description of the edge states for the generic and
M\"{o}bius graphene strip. Here the tight binding Hamiltonian is
used to model the graphene strip with finite width. In
Sec.~\ref{sec:perturb}, a uniform electric field is applied to the
generic and M\"{o}bius graphene strip as a perturbation. We compare
the energy bands of the edge states in the generic and M\"{o}bius
graphene strips under the uniform electric field. Robustness of the
edge states in the M\"{o}bius graphene strip under such perturbation
is demonstrated. And we prove the candidacy of M\"{o}bius graphene
strip as a topological insulator. Additionally, in
Sec.~\ref{sec:gauge}, we discuss the topology-induced non-Abelian
gauge field in the M\"{o}bius graphene strip and its observable
effects in future experiments. Besides, we propose a possible
quantum manipulation mechanism upon the M\"{o}bius graphene strip
through a
magnetic flux. Finally, the conclusion is presented in Sec.~\ref%
{sec:conclusion}.

\section{\label{sec:edge}EDGE STATES IN M\"{O}BIUS GRAPHENE STRIP}

To describe the motion of electrons of graphene, we use the tight binding
model~\cite{Carbon1,Carbon2},
\begin{equation}
H=-J\sum_{\left\langle i,j\right\rangle }[\psi_{a}^{\dagger}(\mathbf{\mathbf{%
r_{i}}}) \psi_{b}(\mathbf{r_{j}})+H.c.], \label{eq:2-1}
\end{equation}
where $\psi_{a}(\mathbf{r_{i}})$($\psi_{b}(\mathbf{r_{i}})$) annihilates an
electron on site $\mathbf{r_{i}}$ of sublattice A(B), and the sum is taken
over the nearest neighbor sites $\left\langle i,j\right\rangle $ with
corresponding hopping constant $J$. For the zigzag graphene strip with
finite width ($N$ regular hexagons in Fig.~\ref{fig1}(a)) along $y$
direction, the $y$ component of the spacial vector $\mathbf{r} $ for A and B
sublattices only take a finite number of values, e.g., $%
y_{m}^{(a)}=[3(N-m)-1]l/2$ for sublattice A, and $y_{m}^{(b)}=[3(N-m)-2]l/2$
for sublattice B. Here, $l$ is the distance between nearest neighbors in the
lattice, and $m=0,1,...,2N-1$. Usually, it is assumed that the length of the
graphene strip ($M$ regular hexagons along x direction, as in Fig.~\ref{fig1}%
(a)) is much larger than the width, namely, $M\gg N$, and periodic boundary
condition is taken as
\begin{equation}
\psi_{a(b)}(\mathbf{r})=\psi_{a(b)}(\mathbf{r}+L\mathbf{e_{x}}),
\label{eq:2-2}
\end{equation}
with $L=\sqrt{3}Ml.$ Thus, it is proper to perform Fourier transformation
only for the $x$ component of $\mathbf{r}$ upon the field operators of
electrons as
\begin{equation}
\psi_{a(b)}(k_{x},y)=\frac{1}{\sqrt{M}}\sum_{x}\psi_{a(b)}(x,y)e^{-ik_{x}x},
\label{eq:2-3}
\end{equation}
and $k_{x}$ can be considered as continuous since $L$ is large enough.

\begin{figure}[ptb]
\begin{centering}
\includegraphics[bb=31 104 474 450,clip,width=3in]{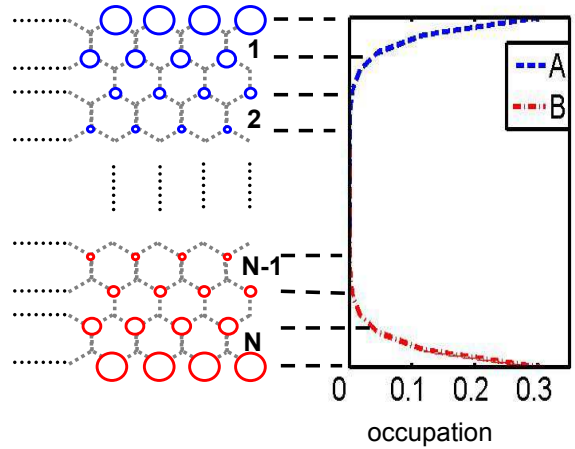}
\par\end{centering}
\par
\centering{}
\caption{ (color online) On the left is the schematic illustration of the
on-site electron occupation of an edge state with $k_{x}=1.2\protect\pi/d$, $%
d=\protect\sqrt{3}l$, in which the size of each circle represents the
magnitude of the electron occupation on that lattice site. And the right
figure displays the corresponding values of electron occupation on
sublattice A (blue circles) and B (red circles) respectively. }
\label{fig2}
\end{figure}

For a zigzag graphene strip, there exist eigenstates strongly
localized on the edges of the strip with their energies lying
exactly within the gap
~\cite{edgestate1,edgestate2,edgestate3,edgestate4}, which are
called edge states. A straightforward calculation gives approximate
edge states
\begin{equation}
\left|\Psi_{\pm}(k_{x})\right\rangle
=D_{e\pm}^{\dagger}(k_{x})\left|vac\right\rangle, \label{eq:2-4}
\end{equation}
which are defined by the corresponding annihilation operators
\begin{equation}
D_{e\pm}(k_{x})=\frac{1}{\sqrt{2}}[A_{e}(k_{x})\pm B_{e}(k_{x})],
\label{eq:2-5}
\end{equation}
where the collective operators
\begin{subequations}
\begin{align}
A_{e}(k_{x})= & \frac{1}{\sqrt{S_{k_{x}}}}\sum_{m=0}^{2N-1}p_{k_{x}}^{m}%
\psi_{a}(k_{x},y_{m}^{(a)}),  \label{eq:2-6-1} \\
B_{e}(k_{x})= & \frac{1}{\sqrt{S_{k_{x}}}}%
\sum_{m=0}^{2N-1}p_{k_{x}}^{2N-1-m}\psi_{b} (k_{x},y_{m}^{(b)})
\label{eq:2-6-2}
\end{align}
respectively represent parts of the edge states localized on the edges, as
shown in Fig.~\ref{fig2}. Here,
\end{subequations}
\begin{equation}
p_{k_{x}}=-2\cos\left(\frac{\sqrt{3}k_{x}l}{2}\right)  \label{eq:2-7}
\end{equation}
represents the decay rate of the single electron on-site occupation
probability along $y$ direction. And
\begin{equation}
S_{k_{x}}=\frac{1-p_{k_{x}}^{4N}}{1-p_{k_{x}}^{2}}  \label{eq:2-8}
\end{equation}
is the normalization constant.
$\left|\Psi_{\pm}(k_{x})\right\rangle$ are the anti-bond and bond
states with respect to the states
$A_{e}^{\dagger}(k_{x})\left|vac\right\rangle$ and
$B_{e}^{\dagger}(k_{x}) \left|vac\right\rangle$. They correspond to
the eigenenergy
\begin{equation}
E_{e\pm}(k_{x})=\pm J\frac{p_{k_{x}}^{2N}(1-p_{k_{x}}^{2})}{1-p_{k_{x}}^{4N}}%
.  \label{eq:2-9}
\end{equation}

Actually, only when $|p_{k_{x}}|<1$ are the edge states defined in
Eq.~(\ref {eq:2-5}) localized at upper and lower edges. The on-site
electron occupation decays exponentially when heading into the bulk,
which is plotted in Fig.~\ref{fig2}. The requirement $|p_{k_{x}}|<1$
means that $k_{x}$ can only be taken between the two neighboring
Dirac points, namely, $k_{x}\in (2\pi /3\sqrt{3}l,4\pi
/3\sqrt{3}l)$. No edge states exist beyond this region.

It is pointed out here that only in the large $N$ limit ($N\rightarrow
\infty $) are the above description for edge states (Eq.~(\ref{eq:2-5}))
accurate. However, for finite $N$, when $k_{x}$ is not in the vicinity of
either Dirac point, namely
\begin{equation}
N(1-|p_{k_{x}}|)\gg 1,  \label{eq:2-10}
\end{equation}%
the above description deviates merely negligibly from the accurate one.

%
\begin{figure}[ptb]
\begin{centering}
\includegraphics[bb=32 207 537 411,clip,width=3in]{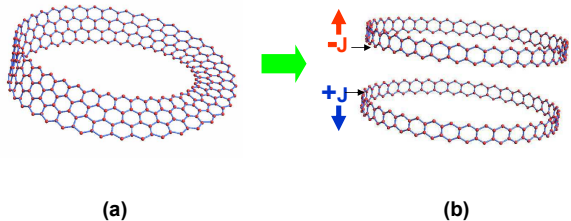}
\par\end{centering}
\caption{ (color online) Schematic illustration of the M\"{o}bius
graphene strip (a) before the unitary transformation $W$ and (b)
after the unitary transformation $W$. After the transformation, the
strip is divided into two independent pseudo strips (red $\uparrow$
and blue $\downarrow$ in (b)) with
no interaction between each other. Besides, the sites on the pseudo edges $%
y=\pm l/2$ obtain extra on-site potential, i.e., +J for the $y=-l/2$ sites
and -J for the $y=l/2$ sites.}
\label{fig3}
\end{figure}


On the other hand, for a M\"{o}bius graphene strip with $N$ regular
hexagons along $y$ direction and $M$ ($M\gg N$) along $x$ direction
(Fig.~\ref{fig1}(b)), the M\"{o}bius boundary condition is
explicitly written as
\begin{equation}
\psi _{a(b)}(\mathbf{r})=\psi
_{b(a)}(\mathbf{\widetilde{r}}+L\mathbf{e_{x}}),
\end{equation}
where $\mathbf{r}=(x,y)$ and $\widetilde{\mathbf{r}}=(x,-y)$, and
$\mathbf{e_{x}}$ is the unit vector along $x$ direction. To make
Fourier transformation on $x$ direction still available, a
position-dependent unitary transformation $W$: $\psi
_{a,b}\rightarrow \psi _{\alpha ,\beta }$
\begin{subequations}
\begin{align}
\psi _{\alpha }(\mathbf{r})=& \frac{1}{\sqrt{2}}[\psi _{a}(\mathbf{r})-\psi
_{b}(\mathbf{\widetilde{r}})]e^{i\frac{\pi }{L}x},  \label{eq:2-11-1} \\
\psi _{\beta }(\mathbf{r})=& \frac{1}{\sqrt{2}}[\psi _{b}(\mathbf{r})-\psi
_{a}(\mathbf{\widetilde{r}})]e^{i\frac{\pi }{L}x}  \label{eq:2-11-2}
\end{align}
for $y>0$, and
\end{subequations}
\begin{subequations}
\begin{align}
\psi _{\alpha }(\mathbf{r})=& \frac{1}{\sqrt{2}}[\psi _{a}(\mathbf{r})+\psi
_{b}(\mathbf{\widetilde{r}})],  \label{eq:2-12-1} \\
\psi _{\beta }(\mathbf{r})=& \frac{1}{\sqrt{2}}[\psi _{b}(\mathbf{r})+\psi
_{a}(\mathbf{\widetilde{r}})]  \label{eq:2-12-2}
\end{align}
for $y<0$ is necessarily used. It can be verified that after the
transformation the new field operators $\psi _{\alpha (\beta )}(\mathbf{r})$
of the electrons satisfy the periodic boundary condition
\end{subequations}
\begin{equation}
\psi _{\alpha (\beta )}(\mathbf{r})=\psi _{\alpha (\beta
)}(\mathbf{r}+L \mathbf{e_{x})}.
\end{equation}%
Then the Hamiltonian of the M\"{o}bius graphene strip becomes
$H=H_{0}+H_{1}$, where
\begin{widetext}
\begin{subequations}
\begin{align}
H_{0}=&
-J{\sum_{<i,j>}}'\left\{\psi_{\alpha}^{\dagger}(\mathbf{r_{i}})\psi_{\beta}(
\mathbf{r_{j}})\exp[i\phi\left(\mathbf{r_{i}},\mathbf{r_{j}}\right)]+H.c.\right\}-J\sum_{x}
[\psi_{\alpha}^{\dagger}(x,-\frac{l}{2})\psi_{\beta}(x,\frac{l}{2})+H.c.],
\label{eq:2-13-1} \\
H_{1}= &
J\sum_{x}[\psi_{\alpha}^{\dagger}(x,-\frac{l}{2})\psi_{\beta}(x,\frac{l}{2})+H.c.
+\psi_{\alpha}^{\dagger}(x,-\frac{l}{2})\psi_{\alpha}(x,-\frac{l}{2})-\psi_{\beta}^{\dagger}(x,\frac{l}{2})
\psi_{\beta}(x,\frac{l}{2})],  \label{eq:2-13-2}
\end{align}
\end{subequations}
\end{widetext}with
\begin{equation}
\phi \left( \mathbf{r_{i}},\mathbf{r_{j}}\right) =\frac{\pi }{L}(\mathbf{%
r_{i}}-\mathbf{r_{j}})\cdot \mathbf{e_{x}}\theta (y_{i}),  \label{eq:2-13-3}
\end{equation}
and $\theta (y)$ is a step function. Here the sum $\sum^{\prime }$ is taken
over all the nearest neighbors except those whose bonds go across the $y=0$
line.

From the above Hamiltonian $H$, we notice that the M\"{o}bius strip
is divided into two separate generic pseudo strips, including $y>0$
strip (lower strip $\downarrow $) and $y<0$ strip (upper strip
$\uparrow $) (see Fig.~\ref{fig3}). There is no coupling between the
upper strip and the lower one, and the $y=\pm l/2$ sites become
{}\textquotedblleft pseudo edges\textquotedblright\ after the
transformation. We should point out that these pseudo edges are not
real in the spacial configuration of the M\"{o}bius strip. Actually,
it follows from Eq.~(\ref{eq:2-13-1}) and Eq.~(\ref {eq:2-13-2})
that there is no coupling between $\psi _{\alpha }(x,-l/2)$ and
$\psi _{\beta }(x,l/2)$ in $H$ while there exist {}\textquotedblleft
on-site potential\textquotedblright\ on the pseudo edges $\pm J\psi
_{\alpha (\beta )}^{\dagger }(x,\mp l/2)\psi _{\alpha (\beta
)}(x,\mp l/2)$. It has been proved that no edge states exist on
these pseudo edges~\cite{spedge1}. In fact, these pseudo edges
appear because of the difference in unitary transformation between
$y>0$ and $y<0$, and possess no topological features of real edges.

As a consequence, there is merely one edge state for each $x$ direction
momentum $k_{x}$ respectively on both upper and lower strips, whose
annihilation operators are defined as
\begin{widetext}
\begin{subequations}
\begin{align}
D_{e\uparrow}(k_{x})= & \frac{1}{\sqrt{S_{k_{x}}}}\sum_{m=N}^{2N-1}[p_{k_{x}}^{2N-1-m}
\psi_{\beta}(k_{x},y_{m}^{(b)})+p_{k_{x}}^{m}\psi_{\alpha}(k_{x},y_{m}^{(a)})],  \label{eq:2-14-1}\\
D_{e\downarrow}(k_{x})= &
\frac{1}{\sqrt{S_{k_{x}}}}\sum_{m=0}^{N-1}[p_{k_{x} -
\frac{\pi}{L}}^{m}\psi_{\alpha}(k_{x},y_{m}^{(a)})-p_{k_{x}-\frac{\pi}{L}}^{2N-1-m}
\psi_{\beta}(k_{x},y_{m}^{(b)})],  \label{eq:2-14-2}
\end{align}
\end{subequations}
\end{widetext}
with corresponding energies
\begin{subequations}
\begin{align}
E_{e\uparrow}(k_{x})= & J\frac{p_{k_{x}}^{2N}(1-p_{k_{x}}^{2})}{%
1-p_{k_{x}}^{4N}},  \label{eq:2-15-1} \\
E_{e\downarrow}(k_{x})= & -E_{e\uparrow}(k_{x}-\frac{\pi}{L}),
\label{eq:2-15-2}
\end{align}
respectively.

Edge states in the graphene strip play an important role in the electron
transport. For the monovalent model of a graphene strip, the Fermi energy is
$E=0$, around which edge states are highly degenerate. Besides, the energy
band of the edge states is only half filled and the graphene strip is a
zero-gap conductor.

\section{\label{sec:perturb}A UNIFORM ELECTRIC FIELD APPLIED AS PERTURBATION}

\begin{figure}[ptb]
\begin{centering}
\includegraphics[bb=6 181 553 639,clip,width=3in]{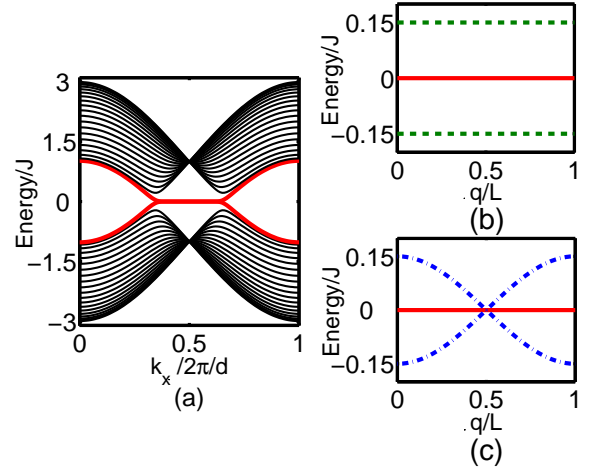}
\par\end{centering}
\caption{ (color online) (a) Energy spectrum of a graphene strip
(M\"{o}bius or generic), whose $(N,M)=(10,100)$, without external
electric field. Obviously the edge states between the two Dirac
points (red line) are approximately degenerate in the energy
spectrum. A schematic illustration of the change in the energy
spectrum is provided in (b) for a generic graphene strip and (c) for
a M\"{o}bius graphene strip, which is caused by the applied uniform
electric field on $y$ direction whose intensity $E$ satisfies
$eEl(3N-1)/2=0.15J$. Here the red solid line represents the energy
level of the edge states without the electric field, while the green
dashed line and the blue dot-dashed line are the changed energy
bands of the edge states in the electric field. Here, as the dual
vector of $k_{x}$ in the inverse Fourier transformation applied only
on the field operators of the edge states, $q$ is analogous to $x$.}
\label{fig4}
\end{figure}


In fact, the conducting behavior of the graphene strip (M\"{o}bius
or generic) under perturbations (e.g., caused by a uniform electric
field), is mainly determined by the properties of its edge states.
For the generic graphene strip, a uniform electric field applied on
$y$ direction will cause a perturbation Hamiltonian
\end{subequations}
\begin{equation}
H_{E}=-eE\sum_{i}y_{i}[\psi_{a}^{\dagger}(\mathbf{r_{i}})\psi_{a}(\mathbf{%
r_{i}}) -\psi_{b}^{\dagger}(\mathbf{\widetilde{r_{i}}})\psi_{b}(\mathbf{%
\widetilde{r_{i}}})],  \label{eq:4-1}
\end{equation}
where $E$ is the electric field intensity. We assume here that the energy
difference introduced by the electric field on opposite edges of the strip
is much smaller than the hopping energy in graphene, namely, $3eENl\ll J$.
Because the energy difference between edge states and bulk states with the
same $k_{x}$ is much larger than the perturbation, transitions between them,
induced by the electric field, can be neglected. Thus we only focus on
transitions between different edge states in the following discussion.

Transition matrix elements between edge states are
\begin{subequations}
\begin{align}
\left\langle \Psi _{\pm }(k_{x})\right\vert H_{E}\left\vert \Psi _{\pm
}(k_{x}^{\prime })\right\rangle =& 0,  \label{eq:4-2-1} \\
\left\langle \Psi _{\pm }(k_{x})\right\vert H_{E}\left\vert \Psi
_{\mp }(k_{x}^{\prime })\right\rangle =& \epsilon _{k_{x}}\delta
(k_{x},k_{x}^{\prime }),  \label{eq:4-2-2}
\end{align}
where $\left\vert \Psi _{\pm }(k_{x})\right\rangle $ are edge states for the
generic graphene strip defined in Eq.~(\ref{eq:2-4}), and
\begin{equation}
\epsilon _{k_{x}}=-eEl\left[ 1+\frac{3}{2}\left( N\frac{1+p_{k_{x}}^{4N}}{%
1-p_{k_{x}}^{4N}}-\frac{1}{1-p_{k_{x}}^{2}}\right) \right] ,  \label{eq:4-3}
\end{equation}%
with $\delta (k_{x},k_{x}^{\prime })$ the Kronecker delta function. Thus the
modified energies of the edge states are
\end{subequations}
\begin{equation}
E_{e\pm }^{\prime }(k_{x})=\pm \sqrt{E_{e+}^{2}(k_{x})+\epsilon
_{k_{x}}^{2}}.  \label{eq:4-4}
\end{equation}

When $N$ is sufficiently large,
\begin{equation*}
\epsilon _{k_{x}}\simeq \epsilon _{\frac{\pi }{\sqrt{3}l}}=-\frac{1}{2}%
eEl(3N-1)
\end{equation*}
is approximately independent of momentum $k_{x}$. Equivalently,
$\epsilon _{k_{x}}$ is approximately a constant in the range of
$k_{x}$ where edge states exist. Besides, when the electric field is
sufficiently strong but $ E\ll J/3eNl$ still holds, the energy
$E_{e\pm }(k_{x})$ of the original edge states approaches zero in
comparison with $|\epsilon _{k_{x}}|$, namely, $|\epsilon
_{k_{x}}|\gg E_{e+}(k_{x})$. In this sense, we approximately obtain
\begin{equation}
E_{e\pm }^{\prime }(k_{x})=\pm \epsilon _{\frac{\pi }{\sqrt{3}l}}.
\label{eq:4-5}
\end{equation}

The above argument means that the originally highly degenerate energy level
of edge states ($E_{e\pm }(k_{x})\simeq 0$) is split into two separate
energy levels by the uniform electric field (see Fig.~\ref{fig4}(b)). This
result could be simply interpreted by the different electric potential on
the upper and lower edges. Since a gap exists between the two energy levels,
the zigzag edges of the graphene strip are no longer conductive.

We notice that the above conclusion is only valid for the case with
generic strips. For a M\"{o}bius graphene strip, due to its inherent
twisted structure, the perturbation Hamiltonian introduced by the
uniform electric field on $y$ direction reads as
\begin{widetext}
\begin{eqnarray}
H_{E}^{(M)} &=&-eE\sum_{i}y_{i}[\psi _{a}^{\dagger }(\mathbf{r_{i}})\psi
_{a}(\mathbf{r_{i}})-\psi _{b}^{\dagger }(\mathbf{\widetilde{r_{i}}})\psi
_{b}(\mathbf{\widetilde{r_{i}}})]\cos \left( \frac{\pi }{L}x_{i}\right)
\notag \\
&=&-\frac{eE}{2}\sum_{k_{x},m=0}^{2N-1}y_{m}^{(a)}\{[\psi _{\alpha
}^{\dagger }(k_{x},y_{m}^{(a)})\psi _{\beta
}(k_{x},-y_{m}^{(a)})+H.c.]+[\psi _{\alpha }^{\dagger
}(k_{x}+\frac{2\pi }{L},y_{m}^{(a)})\psi _{\beta
}(k_{x},-y_{m}^{(a)})+H.c.]\}. \label{eq:4-6}
\end{eqnarray}
\end{widetext}
Here we have assumed a uniform twist for the structure of the strip. It
follows from Eq.~(\ref{eq:4-6}) that the electric field couples the upper
strip with the lower one, and induces transitions between states with
different $k_{x}$, making $k_{x}$ no longer conserved. For the same reason
for a generic graphene strip, we still neglect the transition between edge
states and bulk states. As a consequence, we only need to focus on the
subspace spanned by all the edge states
\begin{equation}
\left\vert \Psi _{\uparrow \left( \downarrow \right)
}(k_{x})\right\rangle =D_{e\uparrow \left( \downarrow \right)
}^{\dagger }(k_{x})\left\vert vac\right\rangle . \label{eq:4-7}
\end{equation}
The transition matrix elements between edge states on the same pseudo strip
vanish, i.e.,
\begin{subequations}
\begin{equation}
\left\langle \Psi _{\uparrow \left( \downarrow \right)
}(k_{x})\right\vert H_{E}^{(M)}\left\vert \Psi _{\uparrow \left(
\downarrow \right) }(k_{x}^{\prime })\right\rangle =0,
\label{eq:4-8-1}
\end{equation}
while the nonzero matrix elements
\begin{widetext}
\begin{equation}
\left\langle \Psi _{\downarrow }(k_{x}^{\prime })\right\vert
H_{E}^{(M)}\left\vert \Psi _{\uparrow }(k_{x})\right\rangle
=\left\langle \Psi _{\uparrow }(k_{x})\right\vert
H_{E}^{(M)}\left\vert \Psi _{\downarrow }(k_{x}^{\prime
})\right\rangle =\epsilon _{k_{x}}^{(M)}[\delta (k_{x},k_{x}^{\prime
})+\delta (k_{x}+\frac{2\pi }{L},k_{x}^{\prime })] \label{eq:4-8-2}
\end{equation}
\end{widetext}
describe transitions between edge states on the upper and lower
pseudo strips. Here $\epsilon _{k_{x}}^{(M)}=\epsilon _{k_{x}}/2$,
and we have neglected the difference between $p_{k_{x}}$ and
$p_{k_{x}+2\pi /L}$. Still, we approximately obtain
\begin{equation*}
\epsilon _{k_{x}}^{(M)}=\epsilon _{\frac{\pi}{\sqrt{3} l}}^{(M)}=-\frac{1}{4}
eEl(3N-1).
\end{equation*}
Since $L$ is large, $k_{x}$ is approximately continuous, and a large
number of $k_{x}$ that satisfy M\"{o}bius boundary condition exist
between the two Dirac points. The conclusion that $E_{e\uparrow
/\downarrow }(k_{x})\simeq 0$ is still valid for the M\"{o}bius
graphene strip when $k_{x}$ is not in the vicinity of either Dirac
point. Thus, the energies of the edge states in the presence of the
uniform electric field are
\end{subequations}
\begin{equation}
E_{q}^{\prime }=\pm 2\epsilon _{\frac{\pi}{\sqrt{3}l}}^{(M)}\cos
\left(\frac{\pi }{L}q\right),  \label{eq:4-9}
\end{equation}
where $q$ is dual vector of $k_{x}$ in the inverse Fourier
transformation. It could be concluded that the originally highly
degenerate energy level $ E_{e\uparrow /\downarrow }(k_{x})=0$ is
broadened, by the uniform electric field, into an energy band with
width $4\epsilon _{\pi /\sqrt{3}l}^{(M)}$ (shown in
Fig.~\ref{fig4}(c)). A straightforward explanation for the energy
band broadening is that the electric potential on the edge of the
M\"{o}bius strip varies along the $x$ direction.

%
\begin{figure}[ptb]
\begin{centering}
\includegraphics[bb=45 176 282 381,clip,width=3in]{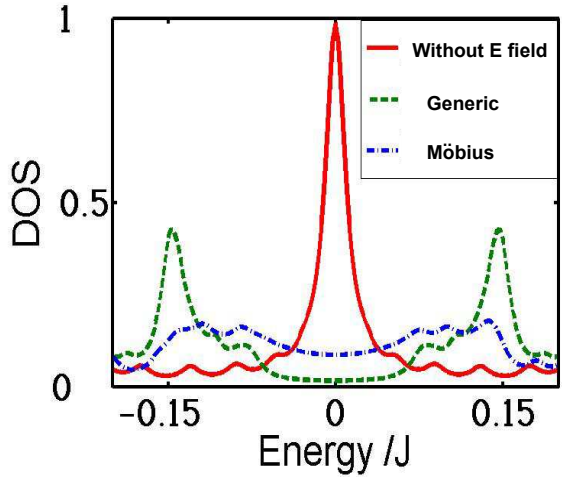}
\par\end{centering}
\caption{ (color online) The density of state (DOS) near $\protect\epsilon=0$
is obtained by deducing the retarded Green's function $G(\protect\epsilon)=(%
\protect\epsilon-H+i\protect\eta)^{-1}$ of the graphene strip with $\protect%
\eta=0.01J$ and $(N,M)=(10,100)$, and determining the spectral function $A(%
\protect\epsilon)=-2$Im$G(\protect\epsilon)$. The black solid line,
the green dashed line and the blue dash-dotted line represent the
DOS of a graphene strip (M\"{o}bius or generic) without electric
field, a generic graphene strip with uniform electric field, and a
M\"{o}bius graphene strip with uniform electric field,
respectively.} \label{fig5}
\end{figure}


It is important to point out that there exist no energy gap for a
M\"{o}bius graphene strip. The zigzag edge is still conductive even
in presence of an external electric field. A M\"{o}bius graphene
strip with such typical feature is referred to as a topological
insulator. Numerical calculations for the density of states (DOS)
near $\epsilon =0$ (Fig.~\ref{fig5}) obviously support the above
theoretical predictions. The DOS-energy curve explicitly
demonstrates the existence of edge states in the vicinity of
$\epsilon =0$ when no external electric field is applied. In this
case, for both the generic and M\"{o}bius graphene strips, there is
only one peak centered exactly at $\epsilon =0$, indicating high
degeneracy of edge states. When uniform electric field is applied on
a generic graphene strip, the peak turns to two peaks centered at
$\pm \epsilon _{\pi /\sqrt{3}l}$, respectively, meaning that the
highly degenerate energy level is split by the electric field. In
this case, the DOS at $\epsilon =0$ approximately equals to zero,
indicating the existence of a gap. Contrarily, with electric field
applied on a M\"{o}bius graphene strip, the DOS has apparently
nonzero value for any $\epsilon $ between $\pm \epsilon _{\pi
/\sqrt{3}l}$. This numerical result agrees well with our above
analytical prediction about the energy spectrum of generic and
M\"{o}bius strips with and without external electric field, and
indicates that the M\"{o}bius graphene strip is a topological
insulator, but a generic strip is not.

\section{\label{sec:gauge}NON-ABELIAN GAUGE FIELD AND OBSERVABLE EFFECTS IN
TRANSMISSION}

%
\begin{figure}[ptb]
\begin{centering}
\includegraphics[bb=118 248 512 476,clip,width=3in]{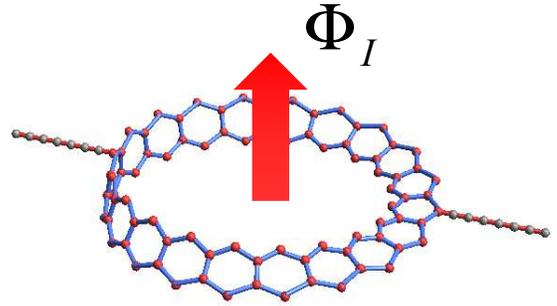}
\par\end{centering}
\par
\centering{} \caption{ (color online) Schematic illustration of a
M\"{o}bius graphene ring with two leads connected to its opposite
sides and an external magnetic flux $\Phi_{I}$ thread its center.}
\label{fig6}
\end{figure}


In this section, we study the transportation properties of electrons
in the M\"{o}bius graphene strip to demonstrate the topological
effects similar to that for M\"{o}bius ladders~\cite{Moladder1}. By
comparing the Hamiltonian Eq.~(\ref{eq:2-13-1}) of a M\"{o}bius
graphene strip with that of a generic graphene strip, it is
recognized that, in the M\"{o}bius graphene strip, there is a phase
shift on the hopping constant for the nearest neighbors
$\left\langle i,j\right\rangle$ whose relative vectors
$\mathbf{r_{i}}-\mathbf{r_{j}}$ have nonzero $x$ component. Similar
to the case of the M\"{o}bius ladder, this phase shift in
Eq.~(\ref{eq:2-13-3}) can be described in terms of a non-Abelian
gauge field
\begin{equation*}
\mathbf{A}(\mathbf{r})=(A_{x},A_{y})=(\frac{c\hbar \pi}
{eL},0)\theta (y)
\end{equation*}
in the continuous limit. The presence of this gauge field changes
the canonical momentum $\mathbf{p}$ into
$\mathbf{p}+e\mathbf{A}(\mathbf{r})$. The gauge field only exists in
the lower strip, to result in an effective magnetic flux $\Phi
=ch/2e$ along the positive $y$ direction thread the center of the
lower strip, which is bent into a ring. Thus when an electron
travels one round along the lower strip, such a gauge field may
bring about a phase shift $\pi $. This effect could be
experimentally testable when leads are connected on opposite sides
of the strip (see Fig.~\ref{fig6}). Then, there would be no
transmission between two leads through the lower strip, due to the
destructive interference between electrons passing through two
possible paths from one lead to the other. Such destructive
interference is analogous to that in the usual Aharonov-Bohm
effect~\cite{ABeffect1,ABeffect2}. However, this phenomenon in our
system is totally induced by non-trivial topology, instead of a real
magnetic flux.

To demonstrate the non-Abelian nature of the gauge field more clearly, we
only consider the special graphene strip with only one regular hexagon along
$y$ direction ($N=1$), which is simply an aromatic hydrocarbon chain (see
Fig.~\ref{fig6}). There are $4$ possible values of $y$ in the lattice of
this strip, $y=\pm l/2$ and $y=\pm l$ respectively. The spinor
representation of electrons in a regular hexagon lattice is
\begin{equation}
\Psi (x)=\left(
\begin{array}{c}
\psi _{a}(x) \\
\psi _{b}(x) \\
\psi _{a}^{\prime }(x) \\
\psi _{b}^{\prime }(x)
\end{array}
\right) \equiv \left(
\begin{array}{c}
\psi _{a}(x+\frac{\sqrt{3}}{2}l,l) \\
\psi _{b}(x,\frac{l}{2}) \\
\psi _{a}(x,-\frac{l}{2}) \\
\psi _{b}(x+\frac{\sqrt{3}}{2}l,-l)%
\end{array}
\right) ,  \label{eq:3-1}
\end{equation}
where we have neglected differences in the $x$ coordinates between field
operators and the $y$ coordinates have been omitted. Performing the unitary
transformation $W$ on $\Psi (x)$ , we have
\begin{equation}
\widetilde{\Psi }(x)=W\Psi (x)=\frac{1}{\sqrt{2}}\left(
\begin{array}{c}
(\psi _{a}(x)-\psi _{b}^{\prime }(x))e^{i\frac{\pi }{L}x} \\
(\psi _{b}(x)-\psi _{a}^{\prime }(x))e^{i\frac{\pi }{L}x} \\
\psi _{a}^{\prime }(x)+\psi _{b}(x) \\
\psi _{b}^{\prime }(x)+\psi _{a}(x)%
\end{array}
\right) .  \label{eq:3-2}
\end{equation}
It can be verified that $\widetilde{\Psi }(x)$ is a periodic
function since $\widetilde{\Psi }(x+L)=\widetilde{\Psi }(x)$. Then
the Hamiltonian of the M\"{o}bius graphene strip becomes
\begin{eqnarray}
H &=&-J\sum_{x}{\widetilde{\Psi }^{\dagger }(x)U\widetilde{\Psi
}(x)+[ \widetilde{\Psi }^{\dagger }(x)T\widetilde{\Psi
}(x+\sqrt{3}l)+H.c.]}  \notag\\
&=&-J\sum_{k_{x}}\widetilde{\Psi }^{\dagger
}(k_{x})[U+(Te^{i\sqrt{3}k_{x}l}+H.c.)]\widetilde{\Psi }(k_{x}),
\label{eq:3-3}
\end{eqnarray}
where
\begin{eqnarray*}
U &=&\left(
\begin{array}{cccc}
0 & 1 & 0 & 0 \\
1 & -1 & 0 & 0 \\
0 & 0 & 1 & 1 \\
0 & 0 & 1 & 0%
\end{array}%
\right) , \\
T &=&\left(
\begin{array}{cccc}
0 & e^{-i\frac{\pi }{M}} & 0 & 0 \\
0 & 0 & 0 & 0 \\
0 & 0 & 0 & 0 \\
0 & 0 & 1 & 0%
\end{array}%
\right) ,
\end{eqnarray*}%
and
\begin{equation*}
\widetilde{\Psi }(k_{x})=\frac{1}{\sqrt{M}}\sum_{x}\widetilde{\Psi }%
(x)e^{-ik_{x}x}.
\end{equation*}

In the continuous limit, the above Hamiltonian becomes
\begin{equation}
H=\int_{0}^{2\pi }d\varphi \widetilde{\Psi }^{\dagger }(\varphi )\{v[R(-i%
\frac{\partial }{\partial \varphi })-A_{0}]+R_{0}\}\widetilde{\Psi
}(\varphi ),  \label{eq:3-4}
\end{equation}
where $\varphi =2\pi x/L$, $v=2\pi \sqrt{3}lJ/L$, $R=$ diag$[-\sigma
_{y},\sigma _{y}]$ ($\sigma _{y}$ is the Pauli matrix),
$A_{0}=$diag$[\sigma _{y}/2,0]$, and $R_{0}=J$ diag$[0,1,-1,0]$. We
would like to point out that in the continuous limit, we have
expanded the original Hamiltonian in the $k_{x}$ representation
around $k_{x}=\pi /\sqrt{3}l$. Because $A_{0}$ does not commute with
$R_{0}$, $A_{0}$ is regarded as  a non-Abelian gauge field.
Actually, the observable effect of the non-Abelian gauge field can
be explicitly demonstrated when the strip has small width, since in
this case, a graphene strip may have distinct energy bands instead
of strongly overlapped ones.

%
\begin{figure}[ptb]
\begin{centering}
\includegraphics[bb=14 185 584 631,clip,width=3in]{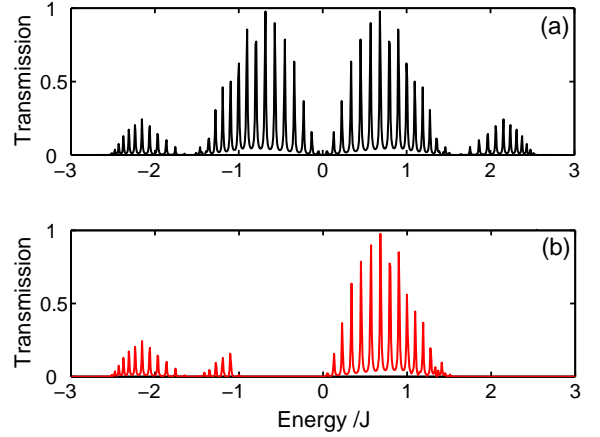}
\par\end{centering}
\par
\centering{} \caption{ (color online) (a) and (b) show the
transmission spectra of a generic graphene strip and a M\"{o}bius
graphene strip, respectively, both of whose parameters
$(N,M)=(1,40)$. Comparing (b) with (a), we find that several peaks
in the transmission spectrum of the generic graphene strip are
missing in that of the M\"{o}bius graphene strip, due to the
existence of a non-Abelian gauge field in the latter one.}
\label{fig7}
\end{figure}


To obtain the transmission spectrum, we calculate the self
energy~\cite{Transport1}
\begin{subequations}
\begin{align}
\Sigma_{L}(k^{\prime }_{x},k_{x}) =&-\frac{J^{\prime }}{2}e^{i\sqrt{3}%
k^{\prime }_{x}l}\left(
\begin{array}{cccc}
1 & 0 & 0 & 0 \\
0 & 0 & 0 & 0 \\
0 & 0 & 0 & 0 \\
0 & 0 & 0 & 1%
\end{array}%
\right), \\
\Sigma_{R}(k^{\prime }_{x},k_{x}) =&-\frac{J^{\prime }}{2}e^{i(\sqrt{3}%
k^{\prime }_{x}l+\frac{1}{2}k_{x}L)}\left(
\begin{array}{cccc}
1 & 0 & 0 & 0 \\
0 & 0 & 0 & 0 \\
0 & 0 & 0 & 0 \\
0 & 0 & 0 & 1%
\end{array}%
\right)
\end{align}
\end{subequations}
in the matrix representation (actually the whole matrix is a
$4M\times4M$ one made up of $M^{2}$ above $4\times4$ blocks with
respective $(k^{\prime }_{x},k_{x})$), and $J^{\prime }$ is the
hopping energy in the leads ($ J^{\prime }=1.5J$ in our numerical
calculation). The above self energy results from the connection of
two leads to the strip. Then we obtain the effective retarded Green
function of the graphene strip
\begin{equation}
G(E(k))=\frac{1}{E(k)-H+i\eta- \Sigma_{L}-\Sigma_{R}},
\end{equation}
where $E(k)=-2J^{\prime }\cos(\sqrt{3}kl)$ is the energy of
electrons injected through the left lead, and $\eta=0_{+}$. After we
determine the level broadening matrix
$\Gamma_{L,R}=-2$Im$\Sigma_{L,R}$, the transmission coefficient
\begin{equation}
T(E(k))=\mathtt{Tr}[\Gamma_{R}G(E(k))\Gamma_{L}G^{\dag}(E(k))]
\end{equation}
is obtained in a straightforward way. The transmission spectrum of a
generic graphene strip and a M\"{o}bius graphene strip (both width
are $N=1$) are displayed in Fig.~\ref{fig7} (a) and (b),
respectively. It is illustrated by Fig.~\ref{fig7}(b) that incident
electrons with energy $E\in[-J,0]$ are totally reflected by the
M\"{o}bius graphene strip. This numerical result just confirm our
heuristic prediction.

\begin{figure}[ptb]
\begin{centering}
\includegraphics[bb=4 219 585 630,clip,width=3in]{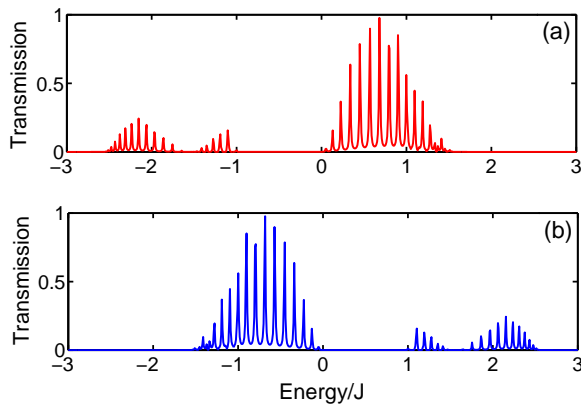}
\par\end{centering}
\par
\centering{} \caption{ (color online) (a) and (b) show the
transmission spectra of a M\"{o} bius graphene ring, whose
parameters $(N,M)=(1,40)$, with an external magnetic flux
$\Phi_{I}=n\Phi_{0}/2$ thread its center, where $n$ is an integer.
(a) corresponds to the $n=$even case, and (b) corresponds to the $n=
$odd case, between which there are obvious differences in the
position of the peaks in the transmission spectrum.} \label{fig8}
\end{figure}

Since the non-Abelian gauge field exists in the lower strip instead
of the upper one, there is a possible way to manipulate the
transmission properties of the M\"{o}bius graphene strip through an
external magnetic flux. As a magnetic flux $\Phi_{I}$ is applied
thread the center of the M\"{o}bius strip (which has the shape of a
ring), a magnetic vector potential appears on both the upper and
lower pseudo strips. When the magnetic flux $\Phi_{I}$ is half of
integer magnetic flux quanta, i.e., $\Phi_{I}=n\Phi_{0}/2$ with
$\Phi_{0}=ch/e$ $(n=1,2,...)$, the total effective magnetic flux in
the lower strip becomes $(n+1)\Phi_{0}/2$ and in the upper strip it
becomes $ n\Phi_{0}/2$. Consequently, when $n$ is even, quantum
transmission is still suppressed in the lower strip, due to the
destructive interference of electrons passing through two possible
paths along the M\"{o}bius ring. Without such destructive
interference, quantum transmission is allowed in the upper strip in
this case. However, when $n$ is odd, quantum transmission is allowed
in the lower strip and suppressed in the upper one. Based on the
above discussion, changes in the positions of the peaks in the
transmission spectrum are expected to be experimentally observed
when the external magnetic flux changes. Numerical calculations,
illustrated in Fig.~\ref{fig8}, clearly verify our above heuristic
discussions. Therefore, this magnetic-flux-based operation obviously
implements quantum manipulation for electron transport in this
M\"{o}bius nanostructure.

%


Next, we further consider the quantum transport of electrons in the low
energy excitation regime. Actually, these electrons locate in the energy
band of the edge states. The energies of edge states in both the lower and
upper strips (see Fig.~\ref{fig3}) are close to the Fermi level of the
system. Here, the energies of the lower edge states are below the Fermi
level, thus these edge states are all occupied at zero temperature.
Contrarily, the upper edge states are not occupied at zero temperature,
since their eigenenergies are above the Fermi level.

\begin{figure}[ptb]
\begin{centering}
\includegraphics[bb=23 85 568 388,clip,width=3.3in]{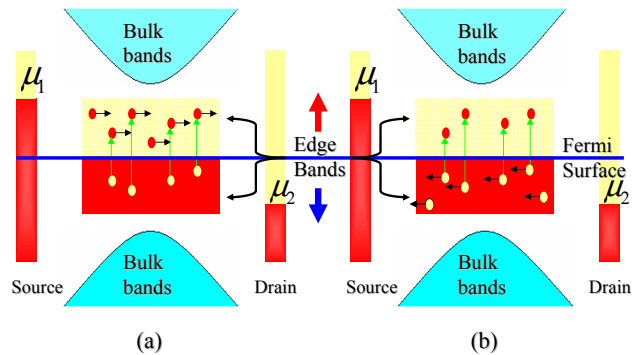}
\par\end{centering}
\par
\centering{} \caption{ (color online) Schematic illustration of
quantum transmission in a M\"{o}bius graphene ring with two leads
connected to its opposite sides and an external magnetic flux
$\Phi_{I}=n\Phi_{0}/2$ thread its center. Here, the up and down
arrows represent the edge states in the upper and lower pseudo
strips of the M\"{o}bius strip (see Fig.~\protect\ref{fig3}). (a)
When $n=$even, the only current carriers are electrons occupying
upper edge states above the Fermi level (red dots). (b) When
$n=$odd, the only current carriers are holes in the energy band of
lower edge states below the Fermi level (light yellow dots).}
\label{fig9}
\end{figure}

When there are an integer number of flux quanta thread the
M\"{o}bius ring ($n$=even in the above discussion), or there exists
no external magnetic flux thread the ring, the fact that quantum
transmission is forbidden in the lower strip means that holes below
the Fermi level cannot be current carriers. Then the only current
carriers in the M\"{o}bius ring between two leads are electrons
occupying the upper edge states (see Fig.~\ref{fig9}(a)).
Oppositely, when the number of flux quanta thread the ring is
half-integer ($n$=odd in the above discussion), quantum transmission
is forbidden in the upper strip. In this case, holes of the lower
strip right below the Fermi level, instead of electrons occupying
the upper edge states, become current carriers between the two leads
(see Fig.~\ref{fig9}(b)). Moreover, if the electron-electron and
electron-phonon interactions could not be ignored in the M\"{o}bius
graphene strip, the transmission rate of electrons and holes would
be different significantly. Thus, accompanying the switch of current
carriers between electrons and holes, transmission rate in the
M\"{o}bius graphene strip may change as well.

%


Finally, we remark on the effect of electric field on the
transmission properties of the M\"{o}bius graphene strip. The
uniform $y$ direction electric field applied to the M\"{o}bius
graphene strip can induce strong coupling between electrons in the
lower and upper strips. In this case, both of the strips can
contribute significantly to quantum transmission. Therefore, the
above forbidden transmission in one of the two pseudo strips no
longer emerges in this case.

\section{\label{sec:conclusion}CONCLUSION}

Oriented by physical realizations of topological insulators and
topological quantum devices, we theoretically studied the electronic
properties of the M\"{o}bius graphene strip, which is an exotic 2D
electron system with a topologically non-trivial edge. Various
properties of the edge states were investigated through the tight
binding model in this paper. We also studied the robustness of edge
states in the M\"{o}bius graphene strip under perturbations, such as
that caused by a uniform electric field. Analytical results about
the exotic natures of such electron system are obtained for the
first time and then confirmed by the following numerical
calculations. Moreover, the physical effects of the non-Abelian
induced gauge structure in the M\"{o}bius graphene strip were
studied by considering the transmission spectrum of the M\"{o}bius
graphene strip with two leads connected to its opposite sides. Based
on such non-Abelian induced gauge field discovered, which is similar
to that for the M\"{o}bius molecular devices studied in
Ref.~\cite{Moladder1}, we proposed a possible manipulation mechanism
for the M\"{o}bius graphene strip through a magnetic flux. In fact,
the robust edge of the M\"{o}bius graphene strip and the non-Abelian
induced gauge field are two characteristic demonstrations of the
nontrivial topology of the M\"{o}bius graphene strip. The above
characters make the M\"{o}bius graphene strip as a candidate for
topological insulators, which may benefit future applications in
quantum coherent devices, and quantum information processing.

\begin{acknowledgements}
One of the authors (Z. L. Guo) thanks T. Shi for helpful
discussions. This work is supported by NSFC No.10474104,
No.60433050, and No.10704023, NFRPC No.2006CB921205 and
2005CB724508.
\end{acknowledgements}

\end{document}